\documentclass{ws-ijmpa}

\begin{document}

\markboth{R.Czy\.zykiewicz -- COSY-11 Collaboration}
{Mechanism of the $\eta$ Meson Production}

%
\catchline{}{}{}{}{}
%

\title{STUDY OF THE PRODUCTION MECHANISM OF THE $\eta$ MESON 
BY MEANS OF ANALYSING POWER MEASUREMENTS}

\author{\footnotesize R.~Czy\.zykiewicz$^{\star,\%}$$^,$\footnote{E-mail 
address: r.czyzykiewicz@fz-juelich.de}~, 
P.~Moskal$^{\star,\%}$, 
H.-H.~Adam$^{\#}$, 
A.~Budzanowski$^{\$}$,
E.~Czerwi\'nski$^{\star}$,
D.~Gil$^{\star}$,
D.~Grzonka$^{\%}$, 
M.~Janusz$^{\star}$, 
L.~Jarczyk$^{\star}$, 
B.~Kamys$^{\star}$, 
A.~Khoukaz$^{\#}$, 
P.~Klaja$^{\star,\%}$, 
B.~Lorentz$^{\%}$, 
J.~Majewski$^{\star,\%}$, 
W.~Oelert$^{\%}$,  
C.~Piskor-Ignatowicz$^{\star}$, 
J.~Przerwa$^{\star,\%}$, 
J.~Ritman$^{\%}$, 
H.~Rohdjess$^{\pm}$, 
T.~Ro\.zek$^{+}$, 
T.~Sefzick$^{\%}$, 
M.~Siemaszko$^{+}$, 
J.~Smyrski$^{\star}$, 
A.~T\"aschner$^{\#}$, 
K.~Ulbrich$^{\pm}$, 
P.~Winter$^{\times}$,  
M.~Wolke$^{\%}$, 
P.~W\"ustner$^{\otimes}$, 
W.~Zipper$^{+}$
}

\address{
$^{\star}$Institute of Physics, Jagellonian University, Cracow, Poland\\ 
$^{\%}$IKP, Forschungszentrum J\"ulich, J\"ulich, Germany\\ 
$^{\#}$IKP, Westf\"alische Wilhelms-Universit\"at, M\"unster, Germany\\
$^{\$}$Institute of Nuclear Physics, Cracow, Poland\\ 
$^{\pm}$Institut f\"ur Strahlen- und Kernphysik, Rheinische 
Friedrich-Wilhelms-Universit\"at, Germany\\
$^{+}$Institute of Physics, University of Silesia, Katowice, Poland\\
$^{\times}$Department of Physics, University of Illinois at 
Urbana-Champaign, Urbana, IL 61801 USA\\
$^{\otimes}$ZEL, Forschungszentrum J\"ulich, J\"ulich, Germany\\
}

\maketitle

\begin{abstract}
Information about the production mechanism of the $\eta$ meson in 
proton-proton collisions can be inferred by confronting the 
experimental studies on the analysing power for the 
$\vec{p}p\to pp\eta$ reaction
with the theoretical predictions of this observable.
The determined analysing powers for Q=10~MeV 
and Q=36~MeV are consistent with zero.
Results show that the predictions of pure pseudoscalar 
meson exchange model fairly describe the experimental data, while 
the predictions of pure vector meson exchange dominance model 
disagree with the data at the level of 4.3~$\sigma$.

\end{abstract}

\section{Introduction}  
After a test measurement of the analysing power for the 
$\vec{p}p\to pp\eta$ reaction at excess energy Q=40~MeV\cite{winter}, 
two additional measurements at Q=10 and Q=36~MeV have been done. 
The experiments have been performed in the Research Center J\"ulich 
(Germany), using the polarised proton beam of the medium-energy storage 
ring COSY\cite{meier} and a H$_2$ cluster target\cite{dombrowski}. 
For the particle detection the COSY-11 detector setup\cite{brauksiepe} 
has been used. In this experiment the missing mass technique has been 
applied in order to identify events in which the $\eta$ mesons have 
been produced.
The goal of this experiment was to determine the main contribution
to the production mechanism of the $\eta$ meson.

\begin{figure}[t]
\begin{center}
\parbox{0.52\textwidth}{
\includegraphics[width=0.48\textwidth]{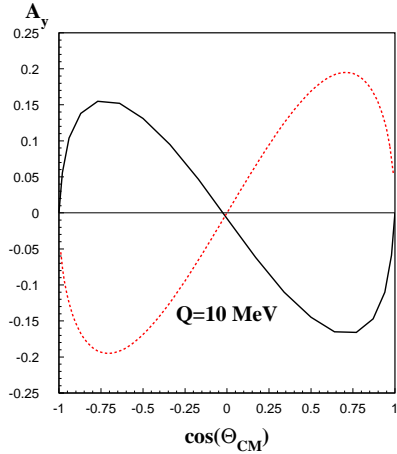}}
\hspace*{1 mm}
\parbox{0.4\textwidth}{\caption{Predictions of the analysing power 
for the $\vec{p}p\to pp\eta$ reaction at Q=10~MeV as a function of 
the center-of-mass (CM) polar angle of the $\eta$ meson emission 
($\Theta_{CM}$). For the explanations of curves see text.}}
\label{fig1}
\vspace*{-4 mm}
\end{center}
\end{figure}

From the near-threshold measurements of the cross sections for the 
$pp\to pp\eta$ reaction\cite{bergdolt:93,tof,moskal-prc}
it was concluded that the production of this meson proceeds via a
two-step process, namely the excitation of the S$_{11}$(1535) resonance
and its subsequent decay into a proton-$\eta$ 
pair\cite{nakayama,hanhart,tord,pmoskal}.
There exist, however, many possibilities leading to the excitation of 
the resonant state of the nucleon in this reaction. The simplest way 
to answer the question whether the excitation of the S$_{11}$(1535) 
proceeds via the exchange of the pseudoscalar- or vector mesons (or 
exchange of both types of mesons) is the measurement of the analysing power
for the $\vec{p}p\to pp\eta$ reaction. In Figure~\ref{fig1}, predictions for
two scenarious mentioned above are depicted for the excess energy
Q=10~MeV.

The solid line in Fig.~\ref{fig1} is a prediction for the pseudoscalar 
meson exchange model\cite{nakayama,nakayama2}, where the excitation of 
the S$_{11}$(1535) resonance is dominated by the exchange of  
pseudoscalar mesons, whereas the dotted line represents the results of 
the calculations based on vector meson exchange\cite{faldt}.  
Significant differences in the sign and the magnitude of the
analysing power predicted basing upon different assumptions
show the sensitivity of the analysing power to the production mechanism.  

\section{Preliminary Results}
\begin{figure}[h]
\begin{center}
\includegraphics[width=0.48\textwidth]{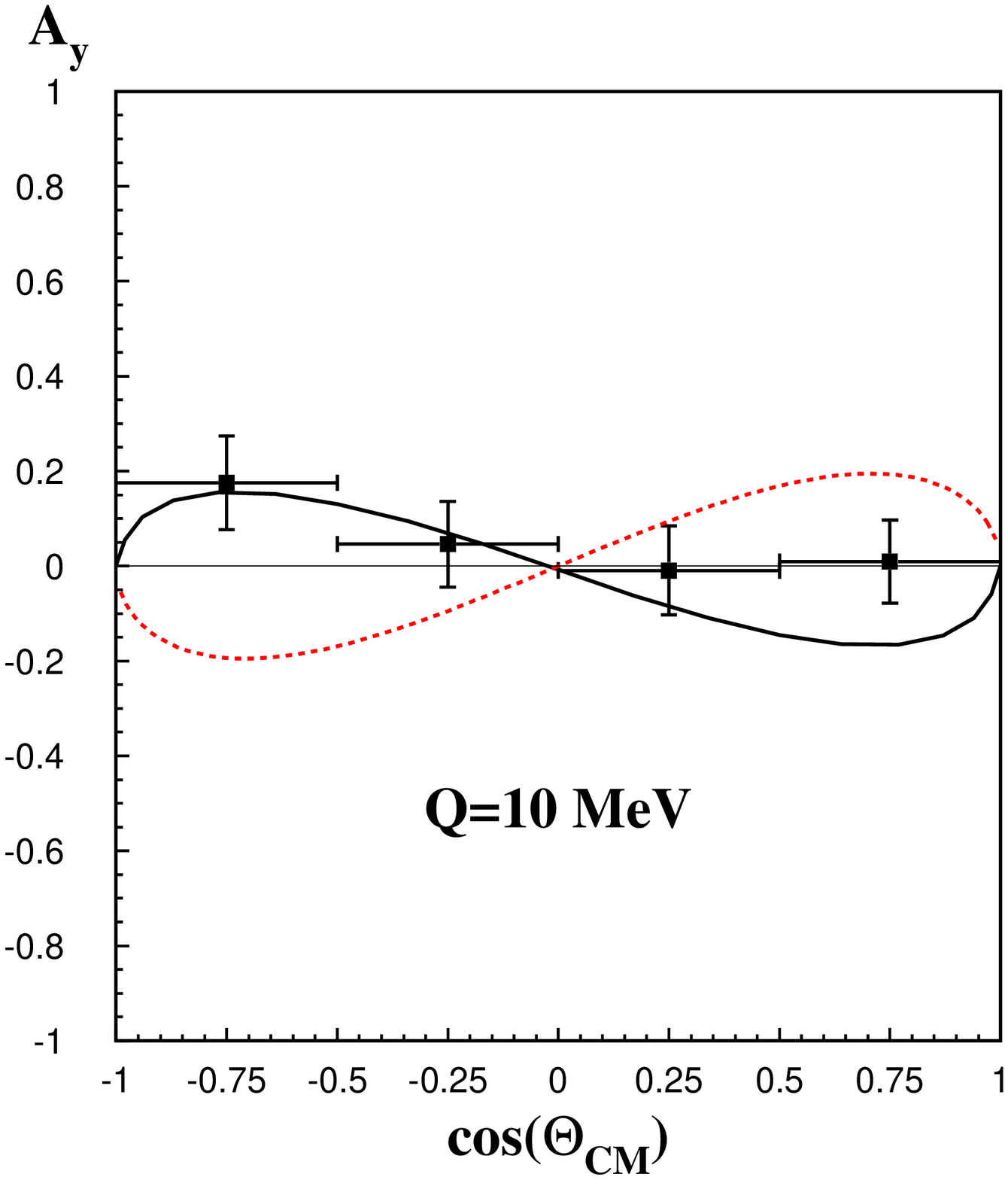}
\includegraphics[width=0.48\textwidth]{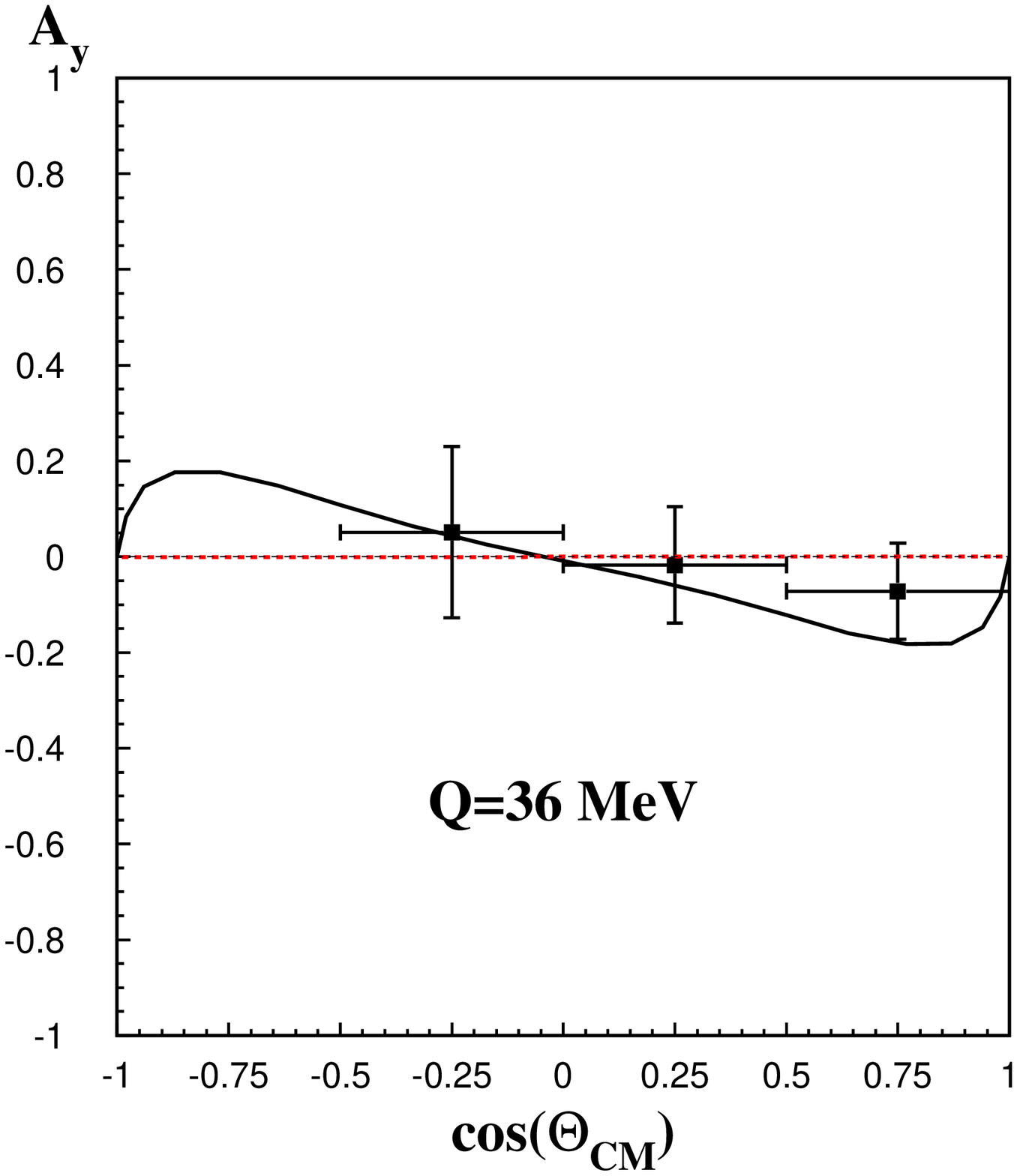}
\caption{{\it Left}: Analysing power function for the 
$\vec{p}p\to pp\eta$ reaction at Q=10~MeV. 
{\it Right}: The same, but for the excess energy Q=36~MeV. 
Vertical bars denote the statistical errors, whereas the horizontal bars 
are to visualize the ranges of averaging. Full lines are predictions 
of the pseudoscalar meson exchange model\protect\cite{nakayama,nakayama2}, 
whereas dotted lines are the predictions of vector meson exchange 
model\protect\cite{faldt}.  For Q=36~MeV prediction of 
vector meson exchange model is consistent with zero. 
The size of the vertical axis has been chosen to cover the 
full range of the allowed values of $A_y$. 
\label{fig2}}
\end{center}
\end{figure}

The experiments have been performed at the beam momenta 
p$_{beam}$=2.010 and 2.085~GeV/c, corresponding to the 
excess energies Q=10 and 36~MeV, respectively\cite{czyzyk1,czyzyk2}. 
Analysing powers have been determined as functions of the center-of-mass 
polar angles of the $\eta$ meson emission -- $\Theta_{CM}$. The preliminary
results of the data analysis are depicted in Fig.~\ref{fig2}.

\begin{figure}[b]
\begin{center}
\parbox{0.52\textwidth}{
\includegraphics[width=0.48\textwidth]{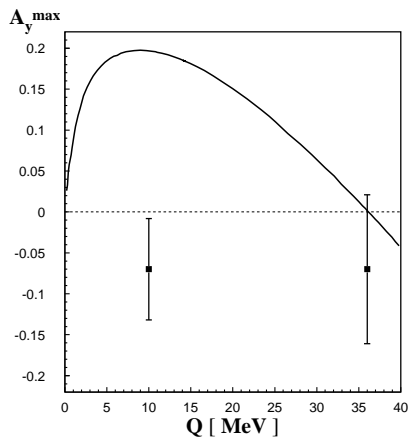}}
\hspace*{3 mm}
\parbox{0.4\textwidth}{\caption{The line denotes the energy 
dependence of A$_y^{max}$ according to the vector meson 
exchange model\protect\cite{faldt}. Two points are the experimentally 
determined values of the A$_y^{max}$ for Q=10 and 36~MeV.}
\label{fig3}}
\end{center}
\end{figure}

In order to verify the correctness of the models
the reduced $\chi^{2}$ for both hypotheses have been calculated.
In calculations the data point for Q=36~MeV and 
$\cos{\Theta_{CM}}=-0.75$ has been excluded due to the low 
statistics and problems with the background evaluation, resulting 
in large systematical error. The reduced value of the $\chi^2$ for the 
pseudoscalar meson exchange model was found to be 
$\chi^{2}_{pseudoscalar}=0.68$, which corresponds to 
the significance level $\alpha_{pseudoscalar}=0.69$, whereas
for the vector meson exchange model $\chi^{2}_{vector}=2.85$ resulting in 
the signi\-ficance level $\alpha_{vector}=0.006$. 

In Ref.~\refcite{faldt} the angular distribution of the 
analysing power is parametrized with the following equation:
\begin{equation}
A_y(\Theta_{CM})=A_y^{max} \sin{2\Theta_{CM}}, 
\label{eq2}
\end{equation}
where the amplitute A$_y^{max}$ is a function of the excess energy Q, 
and its energy dependence is shown as the solid line in Fig.~\ref{fig3}. 

Assuming that the predictions of the vector meson exchange model 
regarding the shape of the angular dependence of the analysing 
power are correct, we have estimated the values of A$_y^{max}$ 
comparing the experimental data with the predicted shape utilizing 
a $\chi^{2}$ test. The values of A$_y^{max}$ for Q=10 and 36~MeV 
have been determined, and are equal to -0.070$\pm$0.062 and 
-0.070$\pm$0.091, respectively. 
Results are presented in Fig.~\ref{fig3}. The predicted A$_y^{max}$ 
function, is at a distance of 4.3~$\sigma$ from the data point 
for Q=10~MeV.

\section{Conclusions}
Our studies indicate that the predictions of the analysing power 
within the pure vector meson exchange models can be 
rejected with significance level of 0.006.
On the other hand, predictions 
of the pseudoscalar meson exchange model, where $\pi$ meson is the dominant meson 
in the internuclear exchange are in line with the experimental data
with the statistical probability of 69~\%. This suggests that the latter 
scenario is much more probable for the $\eta$ meson production.  
However, the interference in the exchange of both types of mesons 
are not excluded and should be studied theoritically. It is also 
possible that the mesonic or nucleonic currents contribute to the
$\eta$ meson production more than it has been assumed in both models.
Further theoretical analysis of this process is required. 

The analysing powers for Q=10~MeV and Q=36~MeV are consistent with zero, 
within one standard deviation.
This may suggest that the $\eta$ meson is predominantly produced in
the $s$-wave.  This observation is in agreement with the results 
of the analysing power measurements performed by the DISTO 
collaboration\cite{disto} where, interestingly, in the far-from-threshold 
energy region the A$_y$ were found to be consistent with zero within
one standard deviation.

\section{Acknowledgements}
We acknowledge the support of the
European Community-Research Infrastructure Activity
under the FP6 "Structuring the European Research Area" programme
(HadronPhysics, contract number RII3-CT-2004-506078),
of the FFE grants (41266606 and 41266654) from the Research Centre 
J{\"u}lich, of the DAAD Exchange Programme (PPP-Polen),
of the Polish State Committee for Scientific Research
(grant No. PB1060/P03/2004/26),
and of the RII3/CT/2004/506078 - Hadron Physics-Activity - N4:EtaMesonNet.


\begin{thebibliography}{99}
\bibitem{winter} P.~Winter {\it et al.}, 
{\it Phys. Lett.} {\bf B544}, 251 (2002); 
erratum-ibid. {\bf 553} 339 (2003). 

\bibitem{meier} R.~Maier {\it et al.}, 
{\it Nucl. Instr. Meth.} {\bf A390}, 1 (1997).

\bibitem{dombrowski} H.~Dombrowski {\it et al.},
{\it Nucl. Instr. Meth.} {\bf A386}, 228 (1997).

\bibitem{brauksiepe} S.~Brauksiepe {\it et al.}, 
{\it Nucl. Instr.  Meth.} {\bf A376}, 397 (1996);\\
P.~Klaja {\it et al.},
{\it AIP Conf. Proc.} {\bf 796}, 160 (2005);\\
J.~Smyrski {\it et al.},
{\it Nucl. Instr.  Meth.} {\bf A541}, 574 (2005).

\bibitem{bergdolt:93} A.M.~Bergdolt {\it et al.},
{\it Phys. Rev.} {\bf D48}, 2969 (1993);\\  
E.~Chiavassa {\it et al.}, {\it Phys. Lett.} {\bf B322}, 270 (1994);\\ 
H.~Cal\'en {\it et al.}, {\it Phys. Lett.} {\bf B366}, 39 (1996);\\  
H.~Cal\'en {\it et al.}, {\it Phys. Rev. Lett.} {\bf 79}, 2642 (1997);\\
F.~Hibou {\it et al.}, {\it Phys. Lett.} {\bf B438}, 41 (1998);\\
J.~Smyrski {\it et al.}, {\it Phys. Lett.} {\bf B474}, 182 (2000). 

\bibitem{tof} M.~Abdel-Bary {\it et al.}, 
{\it Eur. Phys. J.} {\bf A16}, 127 (2003).

\bibitem{moskal-prc} P.~Moskal {\it et al.},
{\it Phys. Rev.} {\bf C69}, 025203 (2004).

\bibitem{nakayama} K.~Nakayama {\it et al.}, 
{\it Phys. Rev.} {\bf C65}, 045210 (2002).

\bibitem{hanhart} C.~Hanhart 
{\it Phys. Rept.} {\bf 397}, 155 (2004).

\bibitem{tord} G.~F\"aldt, T.~Johansson, C.~Wilkin
{\it Phys.~Scripta} {\bf T99}, 146 (2002). 

\bibitem{pmoskal} P.~Moskal {\it et al.},
{\it Prog. Part. Nucl. Phys.} {\bf 49}, 1 (2002).

\bibitem{czyzyk1} R.~Czy\.zykiewicz {\it et al.}, 
{\it Acta Phys. Slovaca} {\bf 56}, 387 (2006); {\it ArXiv:nucl-ex/0603018}.

\bibitem{czyzyk2} R.~Czy\.zykiewicz 
{\it Schriften des FZ-J\"ulich, Matter and Materials} {\bf 21}, 122 (2004).

\bibitem{nakayama2} K.~Nakayama {\it et al.},
{\it Phys. Rev.} {\bf C68}, 045201 (2003).

\bibitem{faldt} G.~F\"aldt, C.~Wilkin 
{\it Phys. Scripta} {\bf 64}, 427 (2001).

\bibitem{disto} F.~Balestra {\it et al.}, 
{\it Phys. Rev.} {\bf C69}, 064003 (2004).

\end{thebibliography}
\end{document}